\documentclass[twocolumn,amsmath,amssymb,aps]{revtex4}
\usepackage{bm}
\usepackage{graphicx}
\usepackage{dcolumn}
\usepackage{epsfig}
\usepackage[cp1251]{inputenc}
\usepackage[english]{babel}
\usepackage{euscript}
\usepackage{amstext}
\usepackage{amsmath}
\usepackage{amssymb}
\usepackage{subfigure}
\usepackage{wrapfig}
\DeclareGraphicsExtensions{.pdf,.jpg,.eps}

\begin{document}

\preprint{}

\title{Forward Physics with the CMS Experiment at the Large Hadron Collider}

\thanks{Presented at Forward Physics at LHC Workshop (May 27-29, 2010), Elba Island, Italy}

\author{Dmytro Volyanskyy\footnote{on behalf of the CMS collaboration}}
 \email{Dmytro.Volyanskyy@cern.ch}
\affiliation{%
Deutsche Elektronen-Synchrotron DESY \\
Notkestrasse 85, 22607 Hamburg, Germany
}%

\begin{abstract}
The forward physics program of the CMS experiment at the LHC spans a broad range
of diverse physics topics including studies of low-$x$ QCD and diffractive scattering,
multi-parton interactions and underlying event structure,
$\gamma$-mediated processes and luminosity determination,
Monte Carlo tuning and even MSSM Higgs discovery in central exclusive production.
In this article, the forward detector instrumentation around the CMS interaction point
is described and the prospects for diffractive and forward physics using
the CMS forward detectors are summarized. In addition, first observation of forward jets
as well as early measurements of the forward energy flow in the pseudorapidity
range $3.15<|\eta|<4.9$ at $\sqrt{s}=0.9$~TeV, $2.36$~TeV and $7$~TeV are presented.
\begin{description}
\item[PACS numbers:] 
\item[Keywords:] forward physics, diffraction, energy flow
\end{description}

\end{abstract}


\maketitle

\section{The CMS experiment at the LHC}

The Compact Muon Solenoid~(CMS)~[1] is one of two general-purpose particle physics detectors
built at the Large Hadron Collider~(LHC) at CERN. The detector has been designed to study
various aspects of proton-proton ($pp$) collisions at $\sqrt{s}=$14~TeV and heavy-ion(Pb-Pb) collisions
at $\sqrt{s}=$5.5~TeV, that will be provided by the LHC at a design luminosity of $10^{34}~ \rm cm^{-2}s^{-1}$
and of $10^{27}~ \rm cm^{-2}s^{-1}$, correspondingly.
To enhance the physics reach of the experiment the CMS subcomponents must provide high-precision
measurements of the momentum and the energy of collision-products. The CMS detector comprises
the tracking system covering the pseudorapidyty range $-2.5<\eta<2.5$ and the calorimetry system covering
the pseudorapity range $-5<\eta<5$. In addition to that, CMS includes several very forward calorimeters,
whose design and physics potential will be described later in this article.
It should be emphasized that the CMS detector is one of the largest scientific instruments ever built.
It comprises about $76.5$ millions of readout channels in total. 
The detector has been designed, constructed and currently operated by the collaboration consisting
of more than $3500$ scientists from $38$ countries.

First collision data taking at CMS took place in November 2009.
Since then and by the end of May 2010, CMS has collected around $10$~$\rm nb^{-1}$ of collision data.
It should be noted that the quality of collected data is rather good:
more than $99\%$ of CMS readout channels are operational and the CMS data taking efficiency is above $90\%$.
Several tens of $\rm pb^{-1}$ of the $pp$ collision data are expected to be collected by the end of 2010.

\section{Forward detectors around the CMS interaction point}

The maximum possible rapidity at the LHC in $pp$ collisions at $\sqrt{s}=14$~TeV
is $y_{max}=ln~(\sqrt{s}/m_{\pi})\approx 11.5$ and one of the great features of the CMS experiment is that it includes several subdetectors
covering the kinematic region at very small polar angles and so, large values of rapidity.
A schematic view of CMS forward detectors is shown in Figure~1. As can be seen, the CMS forward instrumentation
consists of the Hadronic Forward calorimeter~(HF), the CASTOR and ZDC calorimeters.
All of them are sampling calorimeters. That is, they are made of repeating layers
of a dense absorber and tiles of scintillator. A separate experiment TOTEM as well as proton detectors FP420
are additional forward detectors around the CMS interaction point~(IP5).
They further extend the forward reach available around IP5.
\begin{figure}
\begin{center}
\resizebox{3.2in}{!}{
\rotatebox{0}{
\includegraphics{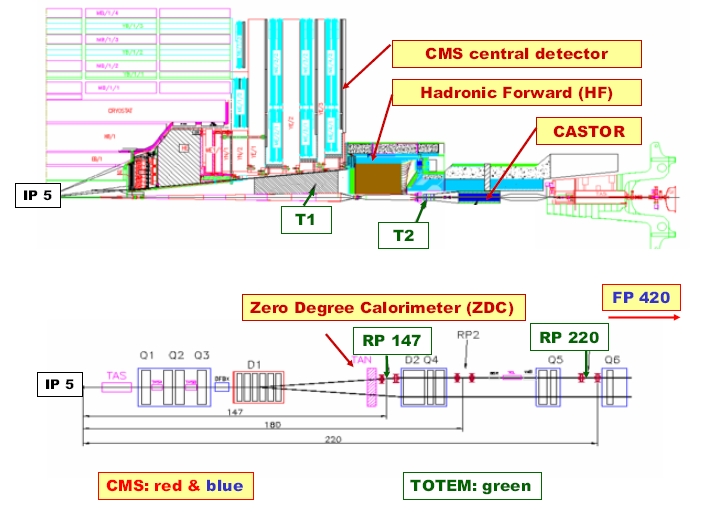}}}
\caption{Layout of the forward detectors around the CMS interaction point.}
\end{center}
\end{figure}
%

\subsection{HF}

The CMS HF detector~[2] includes two calorimeters HF+ and HF--,
which are located at a distance of $11.2$~m on the both sides from the IP5
covering the pseudorapidity range $3<|\eta|<5$. The detector is designed
to carry out the measurements of the forward energy flow and forward jets.
The HF is a Cerenkov sampling calorimeter which uses radiation hard
quartz fibers as the active material and steel plates as the absorber.
The signal in the HF is produced when charged shower particles
pass through the quartz fibers with the energy above the Cerenkov threshold ($190$~keV for electrons).
The generated Cerenkov light is then collected by air-code light guides,
which are connected to photo-multipliers tubes PMTs. The detector fibers run parallel to the beamline
and are bundled to form $0.175 \times 1.175$ ($\Delta\eta \times \Delta\phi$) towers.
Half of the fibers run over the full depth of the absorber,
whereas the other half starts at a depth of $22$~cm from the front of the
detector. These two sets of fibers are read out separately.
Such a structure allows to distinguish showers
generated by electrons and photons, which deposit a large
fraction of their energy in the first $22$~cm, from those generated
by hadrons, which produce signals in both calorimeter
segments. The detector is embedded into a cylindrical steel structure
with the outer radius of $131$~cm and the inner radius of $12.5$~cm
to accommodate the beam pipe. It is azimuthally subdivided into
$20^{0}$ modular wedges, each of which consists of two azimuthal sectors
of $10^{0}$. The detector extends over $10$ interaction lengths and includes
$1200$ towers in total.

\subsection{CASTOR}

The CASTOR (Centauro And STrange Object Reseacrh) detector~[3] is a
quartz-tungsten Cerenkov sampling calorimeter, which is located at a distance of $14.4$~m
from the IP5 and covering the pseudorapidity range $-6.6<\eta<-5.2$.
The tungsten metal has been chosen as the absorber medium in CASTOR,
since it provides the smallest possible shower size. In this detector,
the radiation hard quartz plates used as the active medium are tilted at $45^{0}$
to efficiently capture the Cerenkov light produced by relativistic particles
passing the detector. As in the case of the HF, the produced Cerenkov light is collected
by air-code light guides that are connected to PMTs, which produce signals
proportional to the amount of light collected.
The CASTOR detector is a compact calorimeter with the physical size of about $\rm 65~cm \times 36~cm \times 150~cm$ 
and having no segmentation in $\eta$.  It is embedded into a skeleton, which is made of stainless steel.
The detector consists of $14$ longitudinal modules, each of which comprises $16$ azimuthal sectors
that are mechanically organized in two half calorimeters.
First $2$ longitudinal modules form the electromagnetic section,
while the other $12$ modules form the hadronic section.
In the electromagnetic section, the thicknesses of the tungsten and quartz plates are $5.0$ and $2.0$~mm respectively,
whereas in the hadronic section the corresponding thicknesses are $10.0$ and $4.0$~mm.
With this design, the diameter of the showers of electrons and positrons produced by hadrons
is about one cm, which is an order of magnitude smaller than in other types of calorimeters.  
The detector has a total depth of $10.3$ interaction lengths and includes $224$ readout channels.

\subsection{ZDC}

The CMS ZDC~(Zero Degree Calorimeter) detector~[4] consists of two calorimeters that are located
inside the TAN absorbers at the ends of the straight section of the LHC beampipe at a distance
of 140~m on both sides from the IP5. These are Cerenkov sampling calorimeters with quarz fibers
as the active material and tungsten plates as the absorber material.
The ZDC detector is designed to measure neutrons and very forward photons
providing detection coverage in the pseudorapidity region $|\eta|>8.4$.  
Each ZDC is made up of separate electromagnetic and hadronic sections.
The electromagnetic section consists of $33$ layers of $2$~mm thick tungsten plates
and $33$ layers of $0.7$~mm thick quartz fibers. The hadronic section is made of $24$ layers
of $15.5$~mm thick tungsten plates and $24$ layers of $0.7$~mm thick quartz fibers.
The electromagnetic section is segmented into $5$ horizontal individual readout towers,
whereas the hadronic section is longitudinally segmented into $4$ readout segments.
The tungsten plates are oriented vertically in the electromagnetic section
whereas they are tilted by $45^{0}$ in the hadronic section.
The detector is read out via aircore light guides and PMTs.
It has a total depth of $6.5$ interaction length.
 
\subsection{TOTEM and FP420}

TOTEM~[5] is an independent experiment at the CMS interaction point whose main objectives
are the precise measurement of the total $pp$ cross-section and a study of
elastic and diffractive scattering at the LHC. To achieve optimum forward coverage
for charged particles, TOTEM comprises two tracking telescopes, T1 and T2, that are installed on both sides
from the IP5 in the pseudorapidity region $3.1<|\eta|<6.5$, and Roman Pot stations
that are located at distances of $\pm147$~m and $\pm220$~m from the IP5.
The T1 telescope is located in front of HF and consists of $5$ planes of cathode strip chambers,
while the T2 telescope is located in front of CASTOR and comprises $10$ planes of gas electron multipliers.
For efficient reconstruction of very forward protons, silicon strip detectors are housed in the Roman Pot stations.

FP420~[6] is a proposed detector system, which is supposed to provide
proton detection at a distance of $\pm420$~m from the IP5.
The FP420 detector comprises a silicon tracking system that can be moved transversely
and measure the spatial position of protons, which have been bent out
by the LHC magnets due to the loss of a small fraction of their initial momentum.
The potential physics topics that can be studied
with this detector system include Higgs central exclusive production as well as
a rich QCD and electroweak program.

\section{Physics program}

Extending the physics reach of CMS, the program for forward physics
includes studies of low-$x$ QCD and diffractive scattering,
multi-parton interactions and underlying event structure,
$\gamma$--mediated processes and luminosity determination.
It is also supposed to contribute to the discovery physics
via searches of MSSM Higgs in central exclusive production.

\subsection{Low-x QCD}

A study of QCD processes at a very low parton momentum fraction $x=p_{parton}/p_{hadron}$
is a key to understand the structure of the proton, whose gluon density
is poorly known at very low values of $x$.
Low-$x$ QCD dynamics can be studied in $pp$ collisions
if the parton momentum fraction of one of the colliding protons $x_{1}$
is significantly larger than the parton momentum fraction of the other colliding proton $x_{2}$ ($x_{1}>>x_{2}$).
The result of such a collision is a creation of either jets, prompt-$\gamma$ or Drell-Yan electron pairs
at very low polar angles in the very forward region of the detector.
Low-$x$ QCD studies at CMS will be a continuation of studies of
deep inelastic scattering in electron-proton collisions at HERA,
where low-$x$ QCD dynamics has been explored down to values of $10^{-5}$.
Measurements at HERA have shown that the gluon density
in the proton rises rapidly with decreasing values of $x$.
As long as the densities are not too high this rise can either be
described by the DGLAP model~[7] that assumes strong ordering
in the transverse momentum $k_{T}$ or by the BFKL model~[8] that assumes
strong ordering in $x$ and random walk in $k_{T}$.
Eventually at low enough $x$, the gluon-gluon fusion effects
become important saturating the growth of the parton densities.

At the LHC the minimum accessible $x$ in $pp$ collisions
decreases by a factor of about $10$ for each $2$ units of rapidity.
This implies that a process with a hard scale of $Q \sim 10$~GeV
and within the CASTOR/T2 detector acceptance
can probe quark densities down $x \sim 10^{-6}$.
Such processes include the production of forward jets
and Drell-Yan electron pairs.

\subsubsection{Forward Jets}

A low-$x$ parton distribution function (PDF) of the proton
can be constrained by measuring single inclusive jet cross-section in HF.
Figure~2 illustrates the $log(x_{1,2})$ distribution for parton-parton scattering
in $pp$ collisions at $\sqrt{s}=14$~TeV requiring at least one
jet with the transverse energy above $20$~GeV in the HF acceptance.
As can be seen, by measuring forward jets in HF one can probe $x$ values as low as $10^{-5}$.
\begin{figure}
\begin{center}
\resizebox{3.2in}{!}{
\rotatebox{0}{
\includegraphics{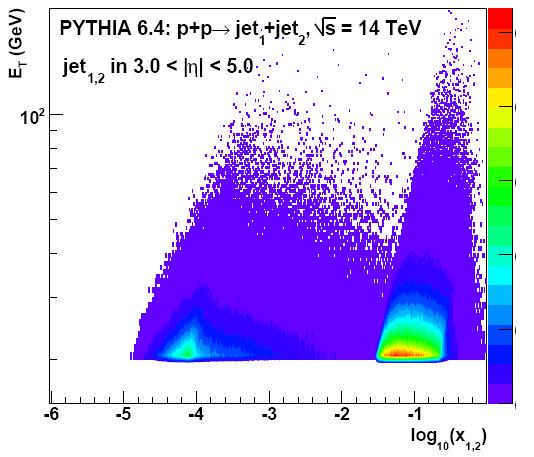}}}
\caption{$log(x_{1,2})$ distribution of two partons producing at least one jet with $E_{T}>20$~GeV in the HF acceptance.}
\end{center}
\end{figure}
A detailed analysis of fully simulated and reconstructed QCD jet events
generated with PYTHIA in the range $p_{T}=20$~GeV/c--$200$~GeV/c in $pp$ collisions
at $\sqrt{s}=14$~TeV  for an integrated luminosity of $1$~$\rm pb^{-1}$ shows that the momentum resolution 
for forward jets in HF is about $18\%$ at $p_{T}=20$~GeV/c and is gradually decreasing to $12\%$ at $p_{T}\geq 100$~GeV/c~[9].

A possibility to gain information on the full QCD evolution to study high order QCD reactions
can be provided by measuring forward jets in the CASTOR calorimeter, that will allow to probe
the parton densities down $10^{-6}$. Apart from that, it has been found that 
a BFKL like simulation
predicts more hard jets in the CASTOR acceptance than the DGLAP model.
Therefore, measurements of forward jets in CASTOR can be used as a good tool
to distinguish between DGLAP and non-DGLAP type of QCD evolution.

Further studies of low-$x$ QCD can be made with Mueller-Navalet dijet events,
which are characterized by two jets with similar $p_{T}$
but large rapidity separation. By measuring Mueller-Navalet dijets
in CASTOR and HF one can probe BFKL-like dynamics and small-$x$ evolution.
\subsubsection{Drell-Yan}
Low-$x$ proton PDFs can also be constructed by measuring electron pairs produced
via the Drell-Yan process $qq\rightarrow\gamma^{*} \rightarrow e^{+}e^{-}$ within the acceptance of CASTOR and TOTEM-T2 station,
whose usage is essential for detecting these events. 
Figure~3 illustrates the distribution of the invariant mass $M$ of the $ee$ system against
the parton momentum fraction $x_{2}$ of one of the quarks, where $x_{2}$ is chosen such that $x_{1} >> x_{2}$.
In this figure, the solid line indicates the kinematic limit,
whereas the region between the dotted lines is the acceptance window for both electrons to be detectable in CASTOR/T2.
The green points show the events with at least one electron lying in CASTOR/T2 acceptance
and the blue points indicate the events with both electrons present within the CASTOR/T2 acceptance,
while the black points correspond to any of the Drell-Yan events generated with PYTHIA.
As can be seen, by measuring two electrons in the CASTOR/T2 acceptance 
one can access $x$ values down to $10^{-6}$ for $M>10$~GeV~[10].
Futhermore, measurements of Drell-Yan events in the CASTOR/T2 acceptance
can be used to study QCD saturation effects. It has been found that
the Drell-Yan production cross section is suppressed roughly by a factor of $2$
when using a PDF with saturation effects compared to one without.
\begin{figure}
\begin{center}
\resizebox{3.2in}{!}{
\rotatebox{0}{
\includegraphics{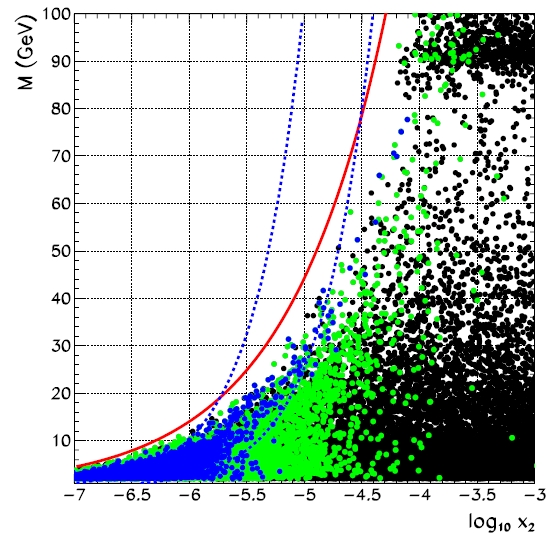}}}
\caption{Acceptance of the CASTOR/T2 detectors for Drell-Yan electrons. See text for details.}
\end{center}
\end{figure}
%

\subsection{Diffraction}
In $pp$ collisions a diffractive process is a reaction $pp \rightarrow X Y$,
where $X$ and $Y$ can either be protons or low-mass systems which may be
a resonance or a continuum state. In all cases, the final states $X$ and $Y$
acquire the energy approximately equal to that of the incoming protons and
carry the quantum numbers of the proton as well as are separated by a Large Rapidity Gap~(LRG).
Diffraction in the presence of a hard scale can be described with perturbative QCD
by the exchange of a colourless state of quarks or gluons, whereas soft diffraction
at high energies is described in the Regge Theory~[11] as a colourless exchange mediated
by the Pomeron having the quantum numbers of the vacuum.
The cross section of hard diffractive processes can be factorized into generalized parton
distributions and diffractive parton distributions functions~(dPDF), which contain
a valuable information about low-$x$ partons. However, the factorization becomes broken
when scattering between spectator partons takes place. This effect is quantified by
the so-called rapidity gap survival probability that can be probed by measuring
the ratio of diffractive to inclusive processes with the same hard scale.
At the Tevatron, the ratio is found to be $O(1\%)$, whereas theoretical expectations
at the LHC vary from a fraction of a percent to up to $30\%$~[12].

The two main types of diffractive processes occurring in $pp$ collisions
are the single diffractive dissociation~(SD) where one of the protons dissociates
and the double diffractive dissociation~(DD) where both protons are scattered into a low-mass system.
The single-diffractive productions of $W$ and dijets are in particular
very interesting processes to study, since they are sensitive
to the quark and gluon content of the PDFs, correspondingly.
They both are hard diffractive processes that can provide information on the rapidity gap survival probability.
A selection of such events can be performed using the multiplicity distributions of tracks
in the central tracker and calorimeter towers in HF plus CASTOR exploiting the fact that diffractive events
on average have lower multiplicity in the central region and in the $"$gap side$"$ than non-difractive ones.
Feasibility studies to detect the SD productions of $W$~[13] and dijets~[14]
have shown that the diffractive events peak in the regions of no activity in HF and CASTOR.

\subsection{Exclusive dilepton production}

Another interesting topic that is going to be studied at CMS is the exclusive dilepton production
$pp \rightarrow ppl^{+}l^{-}$, which can either occur via $\Upsilon$ photoproduction $\gamma p \rightarrow \Upsilon \rightarrow l^{+}l^{-}$
or via the pure QED process $\gamma \gamma \rightarrow l^{+}l^{-}$ that has been observed
by the CDF experiment at the Tevatron~[15]. The latter is an elastic process whose production cross section
is precisely known. As a result, it can potentially serve as an ideal calibration channel
and is going to be used for measuring the luminosity at the LHC.
Using this process an absolute luminosity calibration with the accuracy of $4\%$ is feasible with $100$~$\rm pb^{-1}$ of data~[16].
The dominant background source for this mode is inelastic processes, where one of the proton in the process
does not stay intact but dissociates. It can be significantly suppressed by applying
a veto condition on activity in CASTOR and ZDC.
Exclusive dilepton production occurring via $\Upsilon$ photoproduction is also a mode of interest,
since the cross section of the $\Upsilon$ photoproduction process is sensitive to the generaliszed PDF
for gluons in the proton. Finally, it should be noted that exclusive dimuon production is an ideal alignment channel
for the proposed FP420 proton detectors.
 
\subsection{Multi-parton interactions and forward energy flow}

Multi-parton interactions~(MPI) arise in the region of small-$x$ where parton densities
are large so that the likelihood of more than one parton interaction per event is high.
According to all QCD models, the larger the collision energy the greater the contribution
from multiple parton interactions to the hard scattering process. However, the dependence
of the MPI cross section on the collision energy is not well known and needs to be studied.
A good way to study multiple parton interactions is provided by the energy flow in the forward region,
which is directly sensitive to the amount of parton radiation and MPI.
Measurements of the forward energy flow will allow to discriminate between different MPI models,
which vary quite a lot, and provide additional input to the determination
of the parameters of the existing MPI models. Furthermore, measurements of forward particle production
in $pp$ and Pb-Pb collisions at LHC energies should help to significantly improve
the existing constraints on ultra-high energy cosmic ray models.
The primary energy and composition of the ultra-high energy cosmic rays
are currently determined from Monte Carlo simulations using Regge-Gribov-based approaches~[17] (where the primary particle production
is dominated by forward and soft QCD interactions) with parameters constrained by the existing collider data at the $E_{lab}<10^{15}$~eV,
whereas the measured energies of the ultra-high energy cosmic rays extend up to $10^{20}$~eV and even beyond. 
At the LHC energy of $E_{lab}=10^{17}$~eV, a more reliable determination of the cosmic ray energy and composition becomes possible.
Finally, it should be emphasized that the forward energy flow has never previously been
measured at a hadron collider.

\section{First results from CMS}

\subsection{Observation of forward jets}

A search for forward jets in the pseudorapidity range $3<|\eta|<5$ has been made
as soon as the CMS detector has started to take collision data~[18]. One of the first candidates of a forward dijet
event recorded by CMS at $\sqrt{s}=0.9$~TeV  is shown in Figure~4. The displayed event includes one forward jet and
one backward jet both with a corrected $p_{T}$ above $10$~GeV/c.
\begin{figure}
\begin{center}
\resizebox{3.2in}{!}{
\rotatebox{0}{
\includegraphics{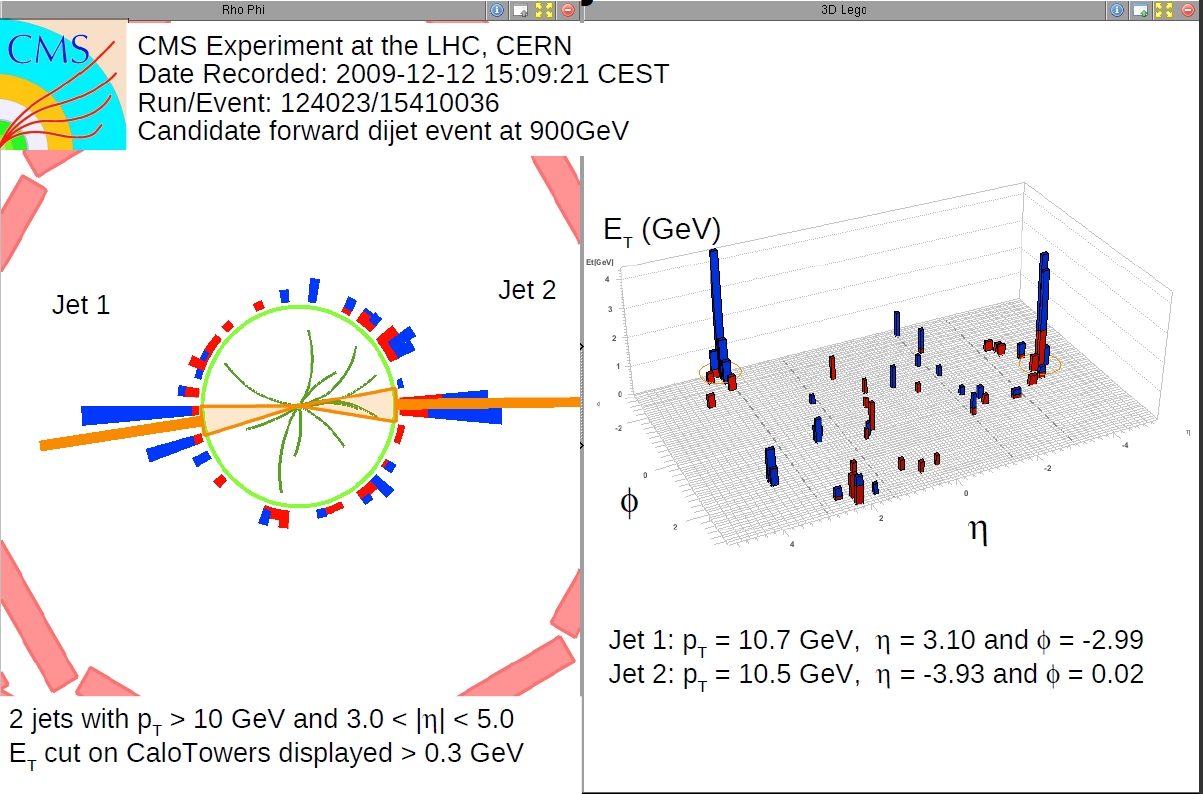}}}
\caption{Display of an event with two forward jets.}
\end{center}
\end{figure}
%

\subsection{Measurement of the forward energy flow}

Early measurements of the energy flow in the forward region of the CMS detector have been made
with minimum bias events using the $pp$ collision data sets collected at $\sqrt{s}=0.9$~TeV and $2.36$~TeV in the fall of 2009
and at $\sqrt{s}=7$~TeV in March 2010~[19]. To select the events of interest the following conditions were imposed.

First, the Beam Scintillator Counters~(BSC) and the Beam Pick-up Timing for the eXperiments~(BPTX),
both are elements of the CMS detector monitoring system, were used to trigger the detector readout.
The BSC devices are located at a distance of $10.86$~m on both sides from the interaction point
covering the pseudorapidity range $3.23<|\eta|<4.65$ and providing hit and coincidence signals
with a time resolution of about $3$~ns. Each BSC comprises $16$ scintillator tiles.
The two BPTX elements are located around the beam pipe at a distance of $\pm175$~m from
the interaction point providing precise information on the bunch structure and timing
of the incoming beam with a time resolution better than $0.2$~ns.
To select the minimum bias events with activity in the forward regions, the coincidence between
a trigger signal in the BSC scintillators and BPTX signals was required for both beams.

Next, to ensure that the selected event is a collision candidate, the events were required
to have at least one primary vertex reconstructed from at least $3$ tracks with a $z$ distance
to the interaction point below $15$~cm and a transverse distance from the $z$-axis smaller than $2$~cm.
Further cuts were applied to reject beam-scrapping and beam-halo events.  Finally, the energy threshold
of $4$~GeV has been imposed to suppress electronic noise in HF.

In this study, the measurement of energy flow has been made at detector level
in the pseudorapidity range $3.15<|\eta|<4.9$ covered by the HF calorimeters.
The energy flow ratio, estimated in this analysis, is defined as
\begin{equation}
 R_{E flow}^{\sqrt{s_{1}}\sqrt{s_{2}}}=\dfrac{ \frac{1}{N_{\sqrt{s_{1}}}} \frac{dE_{\sqrt{s_{1}}}}{d\eta}  } {\frac{1}{N_{\sqrt{s_{2}}}}  \frac{dE_{\sqrt{s_{2}}}}{d\eta}} \;\;\;,
\label{eq:eq}
\end{equation}
where $N_{\sqrt{s}}$ is the number of selected events, $dE_{\sqrt{s}}$  is the energy deposition integrated over $\phi$ in the region $d\eta$,
$\sqrt{s_{1}}$ refers to either $2.36$~TeV or $7$~TeV, whereas $\sqrt{s_{2}}$ refers to $0.9$~TeV.
The pseudorapidity range is divided into five bins with a size of $0.35$  in units of $\eta$
following the transverse segmentation of the HF calorimeters.
In Figures~5 and 6, the energy flow ratio is shown for different collision energies
as the average of the HF(+) and HF(--) responses.
\begin{figure}
\begin{center}
\resizebox{3.2in}{!}{
\rotatebox{0}{
\includegraphics{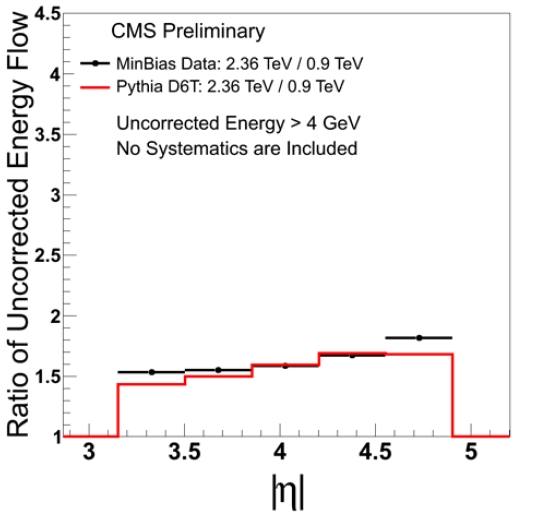}}}
\caption{Energy flow ratio for $\sqrt{s_{1}}=2.36$~TeV to $\sqrt{s_{2}}=0.9$~TeV as a function of $\eta$. See text for details.}
\end{center}
\end{figure}
\begin{figure}
\begin{center}
\resizebox{3.2in}{!}{
\rotatebox{0}{
\includegraphics{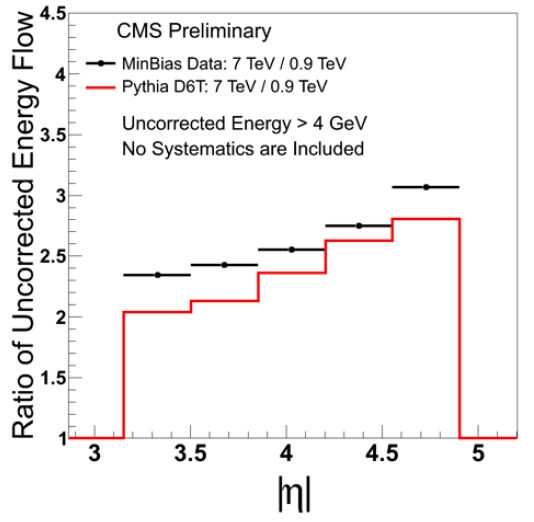}}}
\caption{Energy flow ratio for $\sqrt{s_{1}}=7$~TeV to $\sqrt{s_{2}}=0.9$~TeV as a function of $\eta$. See text for details.}
\end{center}
\end{figure}
In these plots, uncorrected data without systematic uncertainties
are compared to simulated events obtained from PYTHIA tune D6T.
As can be clearly seen, the energy flow gets larger at forward rapidities and with increasing centre-of-mass energy.
Apart from that, it should be noted that the obtained results do approximately agree with the Monte Carlo predictions.
However, no conclusions on the quality of the description can be drawn in this early study due to the missing systematic effects.

\section{Conclusions}

A very rich forward physics program can be made with the CMS detector at the LHC
due to the unprecedented kinematic coverage of the forward region.
All the CMS forward detectors have been successfully commissioned in 2009
and currently take collision data. The first measurement of the forward energy flow
has been performed and forward jets at $|\eta|>3$ have been observed
for the first time at hadron colliders.

\section{Acknowledgments}

I am very grateful to Hannes Jung, Kerstin Borras and many other colleagues working in
the CMS forward physics community for fruitful discussions and kind suggestions.


\end{document}